\input harvmac
\noblackbox
%
%

\font\tiau=cmcsc10
\baselineskip 12pt
\Title{\vbox{\baselineskip12pt \hbox{}
\hbox{} }}
{\vbox{\hbox{\centerline{\bf ELECTRICALLY CHARGED BLACK-HOLES FOR 
THE HETEROTIC STRING}}}}
{\hbox{\centerline{\bf COMPACTIFIED ON A $(10-D)$-TORUS}}}
\vskip3cm
\centerline{\tiau Pablo M. Llatas\foot{llatas@denali.physics.ucsb.edu}}

\vskip.1in
\centerline{\it Department of Physics}
\centerline{\it University of California}
\centerline{\it Santa Barbara, CA 93106-9530}
\vskip .9cm
\noindent{We show that the most general stationary
electrically charged  black-hole solutions of the
heterotic string compactified on a  ($10-D$)-torus (where
$D>3$) can be obtained by using the solution generating
transformations of Sen acting on the Myers and Perry
metric. The conserved charges labeling these black-hole 
solutions are the mass, the angular momentum in
 all allowed commuting planes, and $36-2D$ electric charges. 
General properties of these black-holes are also studied.
}

\Date{May, 1996}

\lref\myp{R.C. Myers and M.J. Perry, Ann. Phys. 172(86)304.}

\lref\car{B. Carter, J. Math. Phys. 10(69)70.}

\lref\sena{A. Sen, Nucl. Phys. B440(95)421, hep-th/9411187.}

\lref\senc{A. Sen, Mod. Phys. Lett. A10(95)2081,  hep-th/9504147.}

\lref\hor{G.T. Horowitz and A. Sen hep-th/9509108}

\lref\horn{J.H. Horne and G.T. Horowitz, Phy. Rev. Lett. D 46(1340).}

\lref\cvea{M. Cvetic and D. Youm, hep-th/9512127.}

\lref\cveb{M. Cvetic and D. Youm, hep-th/9603100.}

\lref\ena{A. Strominger and C. Vafa, hep-th/9601029.}

\lref\enb{C.G. Callan and J.M. Maldacena, hep-th/9602043.}

\lref\enc{G.T. Horowitz and A. Strominger, hep-th/9602051.}

\lref\tend{J.C. Breckenridge, R.C. Myers, A.W. Peet and C. Vafa, 
  hepth/9602065.}

\lref\ene{J.M. Maldacena and A. Strominger, hep-th/9603060.}

\lref\enf{J.C. Breckenridge, D.A. Lowe, R.C. Myers, A.W. Peet, 
A. Strominger and C. Vafa, hep-th/9603078.}

\lref\peet{A. Peet, hep-th/9506200.}

\lref\ms{J. Maharana and J. Schwarz, Nucl. Phys. B390(93)3,
hep-th/9207016.}

\lref\cvep{M. Cvetic and D. Youm, hep-th/9605051.}

\baselineskip 12pt

\newsec{Introduction}

Black-hole solutions emerging in string theory have been extensively studied
in the recent literature. One of the reasons for
such attention is the suggestion that the 
elementary massive string
states (with Planck masses) could be identified with black-holes 
(\refs{\senc} and references therein).
This identification is not necessarily made in the same 
string theory but usually involves dual pictures: 
a classical black-hole solution
of a string theory (``soliton'') is identified with a quantum 
(bound or elementary) state on a dual theory. Identifications of this type 
have been recently employed to provide a statistical derivation
of the Bekenstein-Hawking entropy by identifying the black-holes solutions
with bound configurations of D-branes
(\refs{\ena ,\enb ,\enc ,\tend ,\ene ,\enf}). 

In the present work, we generate the most general rotating electrically
charged black-hole solution of the heterotic string compactified 
on a ($10-D$)-torus (conforming the ``no hair'' theorems). We follow
the work of \refs{\sena}, where the particular case $D=4$ was studied. 
Related works (using the same rotating technique to generate new solutions
from a given one) can be found in \refs{\hor ,\cvea ,\cveb ,\peet }, 
where cases 
of different dimensions, charges and number of rotation planes 
where studied. In the present work, we generate a family of solutions 
depending on a mass, $\Bigl[ {D-1\over 2} \Bigr]$ angular momenta 
($\Bigl[\,\,\Bigr]$
denotes the integer part) and $36-2D$ electrical charges. $16$ electrical
charges come from the $U(1)^{16}$ of the heterotic string 
in ten dimensions (on a general point on the moduli space of
compactifications) and the remaining $2\cdot (10-D)$ electric charges 
come from the compactified 
dimensions (any time we compactify a spatial dimension there
appears generically two new $U(1)$ gauge fields: the Kaluza-Klein $U(1)$ 
field comming
from the metric $G_{\mu\nu}$ and the ``winding'' $U(1)$ field 
originating from the antisymmetric tensor field $B_{\mu\nu}$).
These $1+\bigl[ {D-1\over 2} \bigr] +36-2D$ are the largest 
number of parameters
labeling these stationary black-holes conforming the ``no hair'' 
theorems. In section $2$ we introduce the action for the heterotic
string compactified on a ($10-D$)-torus. The solutions we are about 
to introduce satisfy the equations of motion derived from this action.
In section $3$ we describe the generalization to any $D$ 
of the solution generating technique of \refs{\sena} which generates 
new solutions 
to the equations
of motion from a given one. We will show that the generating boosts
depend on $36-2D$ free parameters, in such a way that if we rotate 
the Myers and Perry metric \refs{\myp} (which already depends on 
$1+\Bigl[ {D-1\over 2}\Bigr]$
parameters corresponding to the mass and angular momenta associated
to rotations in  all commuting planes) we generate the most general
solution we want to construct. 
In section 4 we carry out the rotation and obtain expressions 
for the family of new solutions.
Finally, in section $5$ we study general features 
of these general black-hole solutions, like the mass, electric charges, 
angular momentums, ergosphere and horizons.

\newsec{The Heterotic String Compactified on a ($10-D$)-Torus.}

The low energy effective theory of the ten-dimensional heterotic 
string corresponds to $N=1$ supergravity coupled to $U(1)^{16}$ 
super Yang-Mills (on a generic point of the space of compactification). 
The bosonic content of this 
theory is given by the metric $G_{AB}$, the dilaton $\Phi$, the antisymmetric
tensor $B_{AB}$ and the $U(1)$ gauge bosons $A^I_A$ (where $A,B:0,1,...,9$,
and $I:1,2,..,16$). When we compactify on a ($10-D$)-torus, the metric $G_{AB}$
induces the D-dimensional metric $G_{\mu\nu}$, $10-D$ $U(1)$ 
Kaluza-Klein gauge 
bosons $G_{\mu i}$ and $(11-D)(10-D)/2$ scalars $G_{ij}$ (
here $\mu,\nu:0,1,..D-1$, and $i,j:1,2,..10-D$). The gauge $U(1)$ bosons 
$A^I_A$ originate $16$ $U(1)$ gauge bosons 
$A^I_{\mu}$ and $16\cdot (10-D)$ scalars $A^I_i$. Finally,
the antisymmetric tensor field $B_{AB}$ induces the two form 
$B_{\mu\nu}$, $10-D$ $U(1)$ ``winding'' 
gauge bosons $B_{\mu i}$ and $(10-D)(9-D)/2$ 
scalars $B_{ij}$. Then, the total number of gauge bosons of the compactified 
theory is $36-2D$. These gauge bosons can be arranged in a column 
matrix of vectors 
$A^{(a)}_{\mu}$ (where $a:1,..,36-2D$). The total number of scalars  
(excluding the dilaton) is $260-36D+D^2$.
These scalars can be arranged in a $(36-2D)\times (36-2D)$ matrix 
$M=M^T$ valued on the group G given by:

\eqn\g{
M\subset G, \qquad\qquad G={O(10-D,26-D)\over{O(26-D)\times O(10-D)}}
}
This matrix fulfills $M L M^T =L$ where $L$ is the matrix given by:

\eqn\m{
L=\pmatrix{-I_{26-D} & 0 \cr
        0 & I_{10-D}  \cr
}}
($I_n$ is the $n\times n$ unit matrix).
One can easily check that the dimension of $G$ is precisely $260-36D+D^2$,
in such a way that the scalars of the theory (excluding the dilaton)
 fit in $G$.
In terms of $\Phi$, $M$, $A^{(a)}_{\mu}$, $B_{\mu\nu}$ and $L$, 
the action for 
the heterotic string compactified on a ($10-D$)-torus takes the form
\refs{\ms}:

\eqn\act{
\eqalign{
S=C\int d^D x \sqrt{-G} e^{-\Phi} \bigl[ & R_G +G^{\mu\nu}\partial_{\mu}
  \Phi\partial_{\nu}\Phi +{1\over 8} G^{\mu\nu}Tr(\partial_{\mu} M L
\partial_{\nu} M L)\cr
 &-{1\over 12}G^{\mu\mu '}G^{\nu\nu '}G^{\rho\rho '}
H_{\mu\nu\rho} H_{\mu '\nu '\rho '}\cr
&-G^{\mu\mu '}G^{\nu\nu '}F^{(a)}_{\mu\nu}
(LML)_{ab}F^{(b)}_{\mu '\nu '}\bigr] \cr
}}
where
\eqn\f{
F^{(a)}_{\mu\nu}=\partial_{\mu} A^{(a)}_{\nu}-\partial_{\nu} A^{(a)}_{\mu}
}
and 
\eqn\h{
H_{\mu\nu\rho}=\partial_{\mu}B_{\nu\rho}+2A^{(a)}_{\mu}
L_{ab}F^{(b)}_{\nu\rho}+
{\rm cyclic.}
}

It is straightforward to check that this action is invariant under 
global rotations $\Omega$ leaving the matrix $L$ invariant
($\Omega L{\Omega}^T =L$):

\eqn\om{
M\to \Omega M{\Omega}^T , \qquad A^{(a)}_\mu\to \Omega_{ab}A^{(b)}_{\mu}
}
Under these rotations, $G_{\mu\nu}$, $B_{\mu\nu}$ and $\Phi$ remain
invariant.

\newsec{The Solution-Generating Technique, ``Bar'' Fields.}

Here we describe in some detail the solution-generating technique by 
following the work of Sen \refs{\sena} where the case $D=4$ was studied
(see also \refs{\peet ,\hor ,\cvea ,\cveb}). This technique can be used to, 
given one time-independent solution of the equations of motion of
the action \act , generate a new family of time-independent solutions. 
If we restrict ourselves to time-independent
backgrounds we find two new $U(1)$ gauge fields in the theory: $B_{0m}$ and 
$G_{0m}\over{G_{00}}$. The first $U(1)$ gauge field is related to the
time-independent 
gauge invariance $\delta B_{m0}= \partial_m \lambda_0 -\partial_0 
\lambda_m =\partial_m \lambda_0$. The second gauge field is related to 
invariance under local time-independent translation of the time coordinate. 
One can add these two new $U(1)$ 
gauge fields to our $1\times (36-2D)$ column matrix of vectors 
$A^{(a)}_{\mu}$ and define the $1\times (38-2D)$ matrix of vectors 
$\bar{A}^{(\bar a)}_{\mu}$ (${\bar a}:1,2,..38-2D$) given by:

\eqn\ba{
\eqalign{
{\bar A}^{(a)}_n &\equiv A^{(a)}_n -{G_{0n}\over{G_{00}}} A^{(a)}_0\qquad 
 1\leq a\leq 36-2D\cr
{\bar A}^{(37-2D)}_n &\equiv {1\over 2} {G_{0n}\over{G_{00}}}\cr
{\bar A}^{(38-2D)}_n &\equiv {1\over{2}} B_{0n} +A^{(a)}_0 L_{ab}
  {\bar A}^{(b)}_n \cr
}}
(note that $A^{(a)}_0$ is a scalar under time-independent general 
coordinate transformations). 

Also, the $(36-2D)\times (36-2D)$ $G$-valued matrix
$M$ is promoted to a $(38-2D)\times (38-2D)$ matrix 
$\bar M$. The elements
of the $\bar M$ matrix are given by ($\bar M ={\bar M}^T$):

\eqn\bm{
\eqalign{
{\bar M}_{ab} &\equiv M_{ab}+ 4 {A^{(a)}_0 A^{(b)}_0\over G_{00}}\qquad
 1\leq a,b\leq 36-2D\cr
{\bar M}_{a,37-2D} &\equiv -2 {A^{(a)}_0\over G_{00}}\qquad 1\leq a\leq 
 36-2D\cr
{\bar M}_{37-2D,37-2D} &\equiv {1\over G_{00}}\cr
{\bar M}_{a,38-2D} &\equiv 2(ML)_{ab} A^{(b)}_0 +4{(A^{(b)}_0 L_{bc}
A^{(c)}_0)\over G_{00}} A^{(a)}_0\qquad 1\leq a\leq 36-2D\cr
{\bar M}_{37-2D,38-2D} &\equiv -2 {(A^{(b)}_0 L_{bc} A^{(c)}_0)\over
G_{00}}\cr
{\bar M}_{38-2D,38-2D} &\equiv G_{00}+4A^{(b)}_0 (LML)_{bc} A^{(c)}_0
+4 {(A^{(b)}_0 L_{bc} A^{(c)}_0)^2\over G_{00}}
}}
Also, one defines a ``bar'' metric by:

\eqn\bme{
{\bar G_{nm}}\equiv G_{nm}-{G_{0n}G_{0m}\over G_{00}}
}
and a ``bar'' antisymmetric tensor:
\eqn\bb{
{\bar B_{nm}}\equiv {1\over 2} B_{nm}+{G_{0n}\over G_{00}}(A^{(a)}_m
L_{ab}A^{(b)}_0 -{1\over 2}B_{0m})-(n\leftrightarrow m)
}
Finally, the ``bar'' dilaton is defined through:

\eqn\bd{
{\bar\Phi}=\Phi -{1\over 2}\ln (-G_{00})
}

For time-independent field configurations, the action \act\ can be written
in terms of the ``bar'' fields as:

\eqn\bs{
\eqalign{
S=C\int dt\int d^{D-1} x\sqrt{-\bar G} e^{-\bar\Phi} \bigl[& R_{\bar G} +
{\bar G}^{mn}\partial_m
 {\bar\Phi}\partial_n {\bar\Phi} +{1\over 8} {\bar G}^{mn}
Tr(\partial_m {\bar M} L\partial_n {\bar M} L)\cr
 &-{1\over 12}
{\bar G}^{mm '}{\bar G}^{nn'}{\bar G}^{ll'}
{\bar H}_{mnl} {\bar H}_{m'n'l'}\cr
&-{\bar G}^{mm'}{\bar G}^{nn'}{\bar F}^{({\bar a})}_{mn}
({\bar L}{\bar M}{\bar L})_{{\bar a}{\bar b}}
{\bar F}^{({\bar b})}_{m'n'}\bigr]\cr
}}
where now:
\eqn\blf{
{\bar F}^{({\bar a})}_{mn}=\partial_{m} 
{\bar A}^{({\bar a})}_{n}-\partial_{n} {\bar A}^{({\bar a})}_{m}
}

\eqn\bh{
{\bar H}_{mnl}=\partial_{m}{\bar B}_{nl}+
2{\bar A}^{({\bar a})}_{m}{\bar L}_{{\bar a}{\bar b}}
{\bar F}^{({\bar b})}_{nl}+
{\rm cyclic.}
}
($1\leq {\bar a},{\bar b}\leq 38-2D$) and $\bar L$ is given by:
 
\eqn\lb{
{\bar L} =\pmatrix{L &0 &0\cr
0 &0 &1\cr
0 &1 &0\cr}
}
($L$ is the one in \m ). Again, one straightforwardly verifies that
this action is invariant under the $O(11-D,27-D)$ rotations leaving
the matrix $\bar L$ invariant (${\bar \Omega}{\bar L}{\bar \Omega}^T =
{\bar L}$):

\eqn\rb{
\eqalign{
{\bar M}&\to {\bar M}'={\bar\Omega}{\bar M}{\bar\Omega}^T\cr
{\bar A}^{({\bar a})}_n &\to {{\bar A}'^{({\bar a})}_n} =
{\bar\Omega}_{\bar a\bar b} {\bar A}^{({\bar b})}_n\cr
}}
Once more, $\bar\Phi$, ${\bar G}_{nm}$ and ${\bar B}_{nm}$ remain invariant
under this $O(11-D,27-D)$ rotation.

The spirit of the solution generating technique is the 
following. The rotation \rb\  
in the space of time-independent backgrounds mixes the ``genuine'' 
initial ($36-2D$) $U(1)$ gauge fields $A^{(a)}_{\mu}$ with the 
components $G_{00}$ and $G_{0n}$ of the metric tensor and the 
components $B_{0n}$ of the antisymmetric tensor (as can be seen 
in \ba ). Then, through this mixing one is able
to generate new non-trivial solutions to the equations of motion of \act\
(non-trivial means that we do not produce equivalent solutions). 
Note that due to the fact 
that we do not have magnetic charges (monopoles) the gauge fields 
are well defined globally and so, the new metrics obtained after rotation 
(with whom the original gauge fields mix) are also well defined.

Let us now prove that the non-trivial generating rotations
leaving invariant the asymptotic behaviour of the solution  are 
labeled
by  $36-2D$ free parameters. As we noted above, not all the
$O(11-D,27-D)$  rotations are allowed. $\bar\Omega$ must satisfy
$\bar\Omega \bar L {\bar\Omega}^T =\bar L$. Let us study this condition
(we follow the  lines of \refs{\sena}). The first observation is that
the $U$ matrix given by:

\eqn\u{
U= \pmatrix{I_{36-2D} &0 &0\cr
0 &{1\over\sqrt{2}} &{1\over\sqrt{2}}\cr
0 &{1\over\sqrt{2}} &{-1\over\sqrt{2}}\cr
}}
diagonalizes $\bar L$:

\eqn\bld{
{\bar L}_d \equiv U{\bar L}U^T=\pmatrix{-I_{26-D} &0 &0 &0 \cr
0 &I_{10-D} &0 &0\cr
0 &0 &1 &0\cr
0 &0 &0 &-1\cr
}}
Then, we can write the condition  $\bar\Omega \bar L
{\bar\Omega}^T =\bar L$ as ($U^T U=I_{38-2D}$):

\eqn\obd{
{\bar\Omega}_d {\bar L}_d {\bar\Omega}_d^T={\bar L}_d
}
where we have defined ${\bar\Omega}_d\equiv U {\bar\Omega}U^T$.  
In an asymptotically free space-time (where,
at infinity, $G_{\mu\nu}\to\eta_{\mu
\nu}$ and $A^{(a)}_{\mu} \to 0$), $\bar M$ has the asymptotic form:

\eqn\cote{
{\bar M} \to\pmatrix{I_{36-2D} &0 &0 \cr
   0 & -1 &0\cr
   0 &0 & -1 }
}
(where we take $M=I_{36-D}$). From \bld\ and \cote\ we
conclude  \refs{\sena} that ${\bar\Omega}_d$
(and then, $\bar\Omega$ itself) belongs to a  subgroup $O(1,26-D)\times 
O(10-D,1)$ of $O(11-D,27-D)$. The $O(1,26-D)$ rotates the first 
$26-D$ elements and the element $37-2D$ of a column vector between 
themselves whereas the $O(10-D,1)$  
rotates the elements $27-D,..,36-2D$ and $38-2D$
of a column vector between themselves. However, not all 
the elements of the subgroup 
$O(1,26-D)\times O(10-D,1)$ generate new solutions. We have to quotient
by the global symmetry transformations $\Omega$ in \om\ of the action
\act\  that do not generate new solutions. 
Then, we have to quotient $O(1,26-D)\times O(10-D,1)$ 
by the $O(26-D)$ rotations mixing the first $26-D$ elements of 
a column vector between themselves and the $O(10-D)$ rotations mixing 
the elements $27-D,..,36-2D$ between themselves. Summarizing, 
the rotation group in the $38-2D$ internal space generating inequivalent
solutions of the equations of motion associated to the action \act\ 
 is given by $\bar G$:

\eqn\bg{
\bar G ={O(1,26-D)\times O(10-D,1)\over{O(26-D)\times O(10-D)}}
}
One can easily verify that the dimension of $\bar G$ is precisely 
$36-2D$. 

The strategy to generate the most general electrically
charged black-hole solutions in the heterotic string compactified 
on a ($10-D$)-torus is now clear. First, we take as the generating  solution
the one given by:

\eqn\sol{
\eqalign{
G_{\mu\nu} &= G_{\mu\nu}^{MP}\cr
M &=I_{36-2D}\cr
\Phi = B_{\mu\nu}&=A^{(a)}_{\mu}=0\cr
}} 
where $G_{\mu\nu}^{MP}$ is the metric of Myers and Perry (see next 
section), which from 
now on we will refer as the MP metric. This metric is designed for 
dimensions $D>3$ and we will assume from the rest of the paper 
that this is the case. The MP metric
already depends on $1+\bigl[ {D-1\over 2}\bigr]$ 
parameters: the mass and the angular momentums associated to the maximun
allowed number of rotating planes. $M$ has to be an element of $G$
and $I_{36-2D}$ is the simplest choice. 
\sol\ is trivially a solution of the equations 
of motion of the action \act\ (essentially, the constant 
fields satisfy automatically  
the equations of motion leaving as the only non-trivial equations 
the Einstein equations in vacuum, to which $G^{MP}_{\mu\nu}$ is 
a solution). 

The next step is to use the solution \sol\ to define
the ``bar fields'' \ba\ and \bm . In particular 
${\bar A}^{({\bar a})}_n$ is given by (substitute \sol\ in the 
definitions \ba\ and \bm ):
\eqn\tra{ 
{\bar A}^{({\bar a})}_n ={1\over 2}{G_{0n}\over G_{00}}
{\delta}_{{\bar a},37-2D}
 }
and $\bar M$ is given by:

\eqn\coter{
{\bar M} \to\pmatrix{I_{36-2D} &0 &0 \cr
   0 & (G_{00})^{-1} &0\cr
   0 &0 & G_{00} }
}
 One then performs the ${\bar\Omega}$ 
rotation (depending on $36-2D$ free parameters) and gets
 ${\bar A}'^{({\bar a})}_n$ and ${{\bar M}'}_{{\bar a}{\bar b}}$. 
Finally
one goes backwards in the relations \ba\ and \bm\ and recovers 
the new solutions 
${G'}_{\mu\nu}$, $M'$, $\Phi '$, ${A'}^{(a)}_{\mu}$ and ${B'}_{\mu\nu}$. 
The metric ${G'}_{\mu\nu}$ so obtained depends on $1+\left[ {D-1\over 2}
\right]+36-2D$ parameters, and then is the most general 
electrically charged black-hole 
solution allowed conforming the ``no hair'' theorems.

In the next section we explicitly carry out the $\bar G$ rotation and
find the new solutions.

\newsec{The General Solutions.}

Let us first suppose that we have performed the rotations $\bar\Omega$
in \rb\ and obtained the vector ${\bar A}'^{({\bar a})}_n $ and the 
matrix ${{\bar M}'}_{{\bar a}{\bar b}}$. We can use the relations 
defining the ``bar fields'' in \ba\ and \bm\ to obtain the new solutions
in terms of these ${\bar A}'^{({\bar a})}_n$ and 
${{\bar M}'}_{{\bar a}{\bar b}}$. For 
instance, in \bm\ we see that ${G '}_{00}$ is just given by:

\eqn\gp{
{G '}_{00}={1\over {{\bar M}'}_{37-2D,37-2D}}
}
As another example, we easily obtain ${A '}^{(a)}_0$ from \bm :

\eqn\ap{
{A '}^{(a)}_0 =-{{G'}_{00}\over 2}{{\bar M}'}_{a,37-2D}
= -{{{\bar M}'}_{a,37-2D}\over {2 {{\bar M}'}_{37-2D,37-2D}}}
\qquad 1\leq a\leq 36-2D
}
In a similar way one can compute the expressions for all the fields 
of the new solutions in terms of the elements of the rotated
 vector ${\bar A}'^{({\bar a})}_n $ and the rotated 
matrix ${{\bar M}'}_{{\bar a}{\bar b}}$. Let us collect the result.
The new metric is given by:

\eqn\mep{
\eqalign{
{G'}_{00}&={1\over {{{\bar M}'}_{37-2D,37-2D}}},\qquad\qquad
{G'}_{0n}=2{{\bar A}'^{(37-2D)}_n\over {{{\bar M}'}_{37-2D,37-2D}}}\cr
{G'}_{nm}&=G_{nm}-{G_{0n}G_{0m}\over G_{00}}+
4 {{\bar A}'^{(37-2D)}_n{\bar A}'^{(37-2D)}_m\over 
{{\bar M}'}_{37-2D,37-2D}}\cr
}}
The dilaton is given by:

\eqn\dilp{
{\Phi '}=\Phi -{1\over 2}\ln (G_{00}{{\bar M}'}_{37-2D,37-2D})
}
The expressions for the $U(1)$ gauge fields are:

\eqn\gap{
\eqalign{
{A '}^{(a)}_0 &=-{1\over 2}{{{\bar M}'}_{a,37-2D}\over 
  {{\bar M}'}_{37-2D,37-2D}}\qquad 1\leq a\leq 36-2D\cr 
{A'}^{(a)}_n &= {\bar A}'^{(a)}_n -{{{\bar M}'}_{a,37-2D}\over
  {{\bar M}'}_{37-2D,37-2D}} {\bar A}'^{(37-2D)}_n\cr
}}
The scalar fields of the new solution are given by:

\eqn\escp{
{M'}_{ab}={{\bar M}'}_{ab}-{{{\bar M}'}_{a,37-2D}{{\bar M}'}_{b,37-2D}
 \over {{\bar M}'}_{37-2D,37-2D}}\qquad 1\leq a,b\leq 36-2D
}
And finally, the new antisymmetric tensor is given by:

\eqn\atp{
\eqalign{
{B'}_{0n} &=2{\bar A}'^{(38-2D)}_n +{{{\bar M}'}_{a,37-2D}\over
  {{\bar M}'}_{37-2D,37-2D}}L_{ab}{\bar A}'^{(b)}_n\cr
{B'}_{nm} &={1\over 2}B_{nm}+{G_{0n}\over G_{00}}A^{(a)}_m L_{ab}A^{(b)}_0
  +B_{0n}{G_{0m}\over {2G_{00}}}\cr
 &+{1\over {{\bar M}'}_{37-2D,37-2D}}
  {\bar A}'^{(37-2D)}_n{\bar A}'^{(a)}_m L_{ab}
  {{\bar M}'}_{b,37-2D}+2{\bar A}'^{(37-2D)}_n{\bar A}'^{(38-2D)}_m
  -(n\leftrightarrow m)\cr
}}
(note that this last expression simplifies in 
our case, because we are going to rotate the solution \sol\ where 
$\Phi =A^{(a)}_{\mu}=B_{\mu\nu}=0$).
What remains  now is to compute the rotated 
 vector ${\bar A}'^{({\bar a})}_n $ and matrix 
${{\bar M}'}_{{\bar a}{\bar b}}$ and substitute the result in the 
expressions \mep -\escp\ to obtain the final solution. In \bg\ we saw that 
the generating boosts are elements of the group 
\eqn\gg{
\bar\Omega\in {O(1,26-D)\times O(10-D,1)\over {O(26-D)O(10-D)}}
}
We also saw that $\bar\Omega_d = U\bar\Omega U^T$ is such that
the $O(1,26-D)$ rotates the first $26-D$ elements and the element 
$37-2D$ between themselves whereas the $O(10-D,1)$ rotates the 
elements $27-D,..,36-2D$ and $38-2D$ between themselves. The 
$O(26-D)$ and $O(10-D)$ that we are quotienting out are the rotations
between the first $26-D$ elements and the rotations between the 
elements $27-D,..,36-2D$ respectively. The simplest way to 
parametrize this ${\bar\Omega}_d$ is the following
\refs{\sena}:
\eqn\rot{
{\bar\Omega}_d=U\Omega U^T =R({\bar p},{\bar q})\cdot B(\alpha ,\beta)
}
where $B(\alpha ,\beta)$ performs a boost of angle $\alpha$ between 
the elements $26-D$ and $37-2D$ and a boost of angle $\beta$ 
between the elements $36-2D$ and $38-2D$:
\eqn\Bo{
\eqalign{
B(\alpha,\beta)_{{\bar a},{\bar b}}=
\delta_{{\bar a},{\bar b}}&+
(\cosh{\alpha} -1)(\delta_{{\bar a},26-D}\delta_{{\bar b},26-D} +
   \delta_{{\bar a},37-2D}\delta_{{\bar b},37-2D})\cr
&+\sinh{\alpha}(\delta_{{\bar a},26-D}\delta_{{\bar b},37-2D}
+ \delta_{{\bar a},37-2D}\delta_{{\bar b},26-D})\cr
&+(\cosh{\beta} -1)(\delta_{{\bar a},36-2D}\delta_{{\bar b},36-2D} +
   \delta_{{\bar a},38-2D}\delta_{{\bar b},38-2D})\cr
&+\sinh{\beta}(\delta_{{\bar a},36-2D}\delta_{{\bar b},38-2D}+
  \delta_{{\bar a},38-2D}\delta_{{\bar b},36-2D})\cr
}}
and $R({\bar p},{\bar q})$ is given by $R({\bar p},{\bar q})= 
R_{26-D}({\bar p})\oplus R_{10-D}({\bar q})
\oplus I_2$ where 
$R_N ({\bar k})$ is a $N$-dimensional rotation matrix rotating 
the $N$-dimensional column unit vector $(0,0,..,0,1)$ to the 
$N$-dimensional unit vector $(k_1 ,k_2 ,.., k_N )$:
\eqn\rotk{
\eqalign{
[R_N ({\bar k})]_{ij} =&{\delta}_{ij} \bigl[\bigl(1-{{k_i}^2\over {1+k_N}}
\bigr)(1-{\delta}_{iN})+k_N {\delta}_{iN}\bigr] \cr
&+(1-{\delta}_{ij})\bigl[ ({\delta}_{iN}+{\delta}_{jN}-1)
{k_i k_j\over {1+k_N}} +k_i \delta_{jN} -k_j\delta_{iN}\bigr]\cr
}}
As a consistency check of this parametrization of the generating boosts, 
note that the total number of parameters of the matrix ${\bar\Omega}_d$ 
are $25-D$ coming from the unit vector $\bar p$, $9-D$ coming from the unit 
vector $\bar q$, and the two boost angles $\alpha$ and $\beta$, 
making a total of $36-2D$ parameters (precisely the dimension of the
group \gg ).

Now, we are in position to perform the boost of the  vector 
${{\bar A}}^{({\bar a})}_n $ and the 
matrix ${{\bar M}}_{{\bar a}{\bar b}}$. First, we boost 
${\bar A}'^{({\bar a})}_n $. We need to compute:
\eqn\trab{
{\bar A}'^{({\bar a})}_{\mu} ={\bar\Omega}_{{\bar a}{\bar b}}
{\bar A}^{(\bar b)}_{\mu}=(U^T R({\bar p},{\bar q})B(\alpha ,\beta)
U)_{{\bar a}{\bar b}}{\bar A}^{(\bar b)}_{\mu}
}
From our expressions \u , \Bo , \rotk , and \tra\ one gets the following
result:
\eqn\tap{
\eqalign{
{\bar A}'^{(a)}_n &= {\sinh{\alpha}\over {2\sqrt{2}}} 
  {G_{0n}\over {G_{00}}} p^a \qquad 1\leq a\leq 26-D\cr
{\bar A}'^{(a+26-D)}_n &= {\sinh{\beta}\over {2\sqrt{2}}} 
  {G_{0n}\over {G_{00}}} q^a \qquad 1\leq a\leq 10-D\cr
{\bar A}'^{(37-2D)}_n &= {1\over 4}(\cosh{\alpha} +\cosh{\beta}) 
  {G_{0n}\over {G_{00}}}\cr
{\bar A}'^{(38-2D)}_n &={1\over 4}(\cosh{\alpha} -\cosh{\beta})
  {G_{0n}\over {G_{00}}}\cr
}}

Now, we compute ${\bar M}'$. We need to calculate, in this case:
\eqn\tmp{
{\bar M}'={\bar\Omega}{\bar M}{\bar\Omega}^T =U^T R({\bar p},{\bar q})
B(\alpha ,\beta)U{\bar M}U^T  B^T (\alpha ,\beta)
R^T ({\bar p},{\bar q})U
}
The result is:
\eqn\tmm{
\eqalign{
{{\bar  M}'}_{ab}&=\delta_{ab} +(1+G^{+}){\sinh^2\alpha} p^a p^b 
 \qquad 1\leq a,b\leq 26-D\cr
{{\bar M}'}_{a,b+26-D}&={\sinh\alpha}{\sinh{\beta}}G^{-}p^a q^b\qquad
 1\leq a\leq 26-D,\,\, 1\leq b\leq 10-D\cr
{{\bar M}'}_{a+26-D,b+26-D}&=\delta_{ab}+(1+G^{+}){\sinh^2\beta} q^a q^b
\qquad 1\leq a,b\leq 10-D\cr
{{\bar M}'}_{a,37-2D}&= {a^{+}_1\over G_{00}} p^a\qquad 1\leq a\leq 26-D\cr
{{\bar  M}'}_{a+26-D,37-2D}&={a^{+}_2\over G_{00}} q^a\qquad 
1\leq a\leq 10-D\cr
{{\bar M}'}_{37-2D,37-2D}&={{\Delta}\over G_{00}}\cr
{{\bar M}'}_{a,38-2D}&= {a^{-}_1\over G_{00}} p^a\qquad 1\leq a\leq 26-D\cr
{{\bar  M}'}_{a+26-D,38-2D}&={a^{-}_2\over G_{00}} q^a\qquad 
1\leq a\leq 10-D\cr
{{\bar M}'}_{37-2D,38-2D}&=
{{\bar M}'}_{38-2D,38-2D}={{\delta}\over G_{00}}\cr
}}
where we have defined:
\eqn\tdef{
\eqalign{
G^{\pm}&\equiv {1\over 2}({1\over G_{00}}\pm G_{00})\cr
a^{\pm}_1 &\equiv {\sinh{\alpha}\over 2\sqrt{2}} \bigl[ 
\cosh{\alpha}\pm\cosh{\beta} +2\cosh{\alpha} G_{00}+
 (\cosh{\alpha}\mp\cosh{\beta}){G_{00}}^2\bigr]\cr
a^{\pm}_2 &\equiv {\sinh{\beta}\over 2\sqrt{2}} \bigl[ 
\cosh{\alpha}\pm\cosh{\beta} \pm2\cosh{\beta}G_{00}-
 (\cosh{\alpha}\mp\cosh{\beta}){G_{00}}^2\bigr]\cr  
{\Delta}&\equiv {1\over 4}(\cosh{\alpha}+\cosh{\beta})^2 +
{G_{00}\over 2} ({\sinh^2\alpha} +{\sinh^2\beta})+
{G_{00}^2\over 4}(\cosh{\alpha}-\cosh{\beta})^2\cr
{\delta}&\equiv {1\over 4}({\cosh^2\alpha}-{\cosh^2\beta}) 
(1+{G_{00}}^2)+{1\over 2}({\sinh^2\alpha}-{\sinh^2\beta})G_{00}\cr
}}

Now, we are ready to substitute these results in \gp\ -\atp\ to obtain
the final solution. Here we collect the result. The boosted metric
reads:
\eqn\tr{
\eqalign{
{G'}_{00}&= {G_{00}\over {\Delta}}, \qquad
 {G'}_{0n}={1\over 2}(\cosh{\alpha} +\cosh{\beta}) {G_{0n}\over 
  {\Delta}}\cr
{G'}_{nm}&=G_{nm}+\Bigl[ {(\cosh{\alpha}+\cosh{\beta})^2\over 4{\Delta}}
 -1\Bigr] {G_{0n}G_{0m}\over G_{00}}\cr
}}
The boosted dilaton is:
\eqn\trd{
{\Phi}'=-{1\over 2}\ln{{\Delta}}
}
The $U(1)$ gauge fields are given by:
\eqn\trg{
\eqalign{
{A'}^{(a)}_0 &= -{a^{+}_1 \over 2{\Delta}} p^a \qquad 
 1\leq a\leq 26-D\cr
{A'}^{(a+26-D)}_0 &= -{a^{+}_2 \over 2{\Delta}} q^a \qquad 
 1\leq a\leq 10-D\cr
{A'}^{(a)}_n &={1\over 2}\Bigl[ {\sinh\alpha\over \sqrt{2}} -
{a^{+}_1\over 2\Delta}(\cosh\alpha +\cosh\beta )\Bigr]{G_{0n}\over G_{00}} 
p^a\cr
{A'}^{(a+26-D)}_n &= {1\over 2}\Bigl[ {\sinh\beta\over \sqrt{2}} -
{a^{+}_2\over 2\Delta}(\cosh\alpha +\cosh\beta )\Bigr]{G_{0n}\over G_{00}} 
q^a\cr
}}
The boosted scalars are:
\eqn\trs{
\eqalign{
{M'}_{ab}&={\delta}_{ab} +\Bigl[ (1+G^{+}){\sinh^2\alpha}-{(a^{+}_1)^2\over
{\Delta}G_{00}}\Bigr] p^a p^b\quad 1\leq a,b\leq 26-D\cr
{M'}_{a+26-D,b+26-D}&={\delta}_{ab} +\Bigl[ (1+G^{+}){\sinh^2\beta}-
 {(a^{+}_2)^2\over
{\Delta}G_{00}}\Bigr] q^a q^b\quad 1\leq a,b\leq 10-D\cr
{M'}_{a,b+26-D}&=\Bigl[ G^{-} \sinh{\alpha}\sinh{\beta} -{a^{+}_1 a^{+}_2
\over {\Delta}G_{00}}\Bigr] p^a q^b\quad 1\leq a\leq 26-D ,1\leq b\leq 10-D\cr
}}
Finally, the boosted antisymmetric tensor reads:
\eqn\tan{
\eqalign{
{B'}_{0n}&={1\over 2}\Bigl( \cosh{\alpha} -\cosh{\beta}+{1\over\sqrt{2}}
{a^{+}_2 \sinh{\beta}-a^{+}_1 \sinh{\alpha}\over {\Delta}}\Bigr)
{G_{0n}\over G_{00}}\cr
{B'}_{nm}&=0\cr
}}
At first sight, some of the previous expressions seems to be ill defined
over the ergosphere (i.e., the surface defined by $G_{00}=0$). 
However, a closer inspection shows that all the 
fields are well defined when $G_{00}=0$ (see section 5).

Note that the previous expressions are valid for any metric $G_{\mu\nu}$ 
satisfying the Einstein equations in the vacuum. The only information 
that we have used so far is that (see \sol ) $M=I_{36-2D}$ and 
$\Phi =B_{\mu\nu} =A^{(a)}_{\mu}=0$. Now, to have the most general 
electrically charged black-hole solution of the heterotic string 
compactified on a ($10-D$)-torus we have to substitute the 
Myers and Perry metric 
$G^{MP}_{\mu\nu}$ in the relations \tr -\tan . Here we give the MP 
metric in polar coordinates which are the best adapted to the symmetry of a
rotating black-hole in $\Bigl[ {D-1\over 2}\Bigr]$ commuting planes
(with more than one plane of rotation, the spherical
coordinates lead to complicated, long and asymmetrical expressions). 
On each plane of rotation (there are $\Bigl[ {D-1\over 2}\Bigr]$) we 
select polar coordinates $(r_i ,\theta_i)$ ($i:1,..\Bigl[ {D-1\over 2}
\Bigr]$). When the spatial dimension ($D-1$) is odd, we have to 
introduce another coordinate $z$, labeling the direction in which there 
is not rotation. $a_i$ is the rotation parameter along the plane 
labeled by the coordinates $(r_i ,\theta_i )$. The MP metric 
is given by (remember that $D>3$) \refs{\myp}:
\eqn\tmp{
\eqalign{
G^{MP}_{00}&=h-1,\qquad\qquad G^{MP}_{0{r_i}}=h{\rho r_i\over 
{\rho}^2+{a_i}^2}\cr
G^{MP}_{0{\theta}_i}&=h{a_i {r_i}^2\over {\rho}^2+{a_i}^2},
\qquad\qquad G^{MP}_{0z}={(1-(-1)^{D-1})\over 2}h{z\over \rho}\cr
G^{MP}_{{r_i}{r_j}}&=\delta_{ij} +h{{\rho}^2 r_i r_j\over{ 
({\rho}^2+{a_i}^2)({\rho}^2+{a_j}^2)}}\cr
G^{MP}_{{\theta_i}{r_j}}&=h{\rho a_i {r_i}^2 r_j\over{ 
({\rho}^2+{a_i}^2)({\rho}^2 +{a_j}^2)}},\qquad
G^{MP}_{{r_i}z}={(1-(-1)^{D-1})\over 2}h{z r_i\over 
{\rho}^2+{a_i}^2}\cr
G^{MP}_{{\theta_i}{\theta_j}}&={r_i}^2\delta_{ij} +h{a_i a_j {r_i}^2
 {r_j}^2\over{ 
({\rho}^2+{a_i}^2)({\rho}^2+{a_j}^2)}}\cr
G^{MP}_{{\theta_i}z}&={(1-(-1)^{D-1})\over 2}h{a_i {r_i}^2 z\over{ 
\rho ({\rho}^2+{a_i}^2)}},\qquad G^{MP}_{zz}={(1-(-1)^{D-1})\over 2}\Bigl( 1+
h {z^2\over {\rho}^2}\Bigr)\cr
}}
In the previous expressions, $\rho =\rho(r_i ,a_i)$ ($i:1,.. \Bigl[
{D-1\over 2}\Bigr]$) is the function 
defined by the constraint:
\eqn\trho{
\sum_i \Bigl( {{r_i}^2\over {{\rho}^2+{a_i}^2}}\Bigr) +{(1-(-1)^{D-1})
\over 2} {z^2\over {\rho}^2}=1
}
and $h$ is given by:
\eqn\th{
h= {(1+(-1)^{D-1})\over 2}{\mu {\rho}^2\over {\Pi F}}+
{(1-(-1)^{D-1})\over 2}{\mu \rho\over {\Pi F}}
}
being:
\eqn\tb{
\Pi =\prod_i ({\rho}^2 +{a_i}^2),\qquad\qquad F=1-\sum_i {{a_i}^2 
{r_i}^2\over {({\rho}^2 +{a_i}^2)^2}}
}
An important consequence of the constraint \trho\ (useful when computing
the conserved charges at infinity) is that:
\eqn\limt{
\lim_{r_{i_1},\cdot\cdot ,{r_{i_L}}\to\infty}\rho^2=\sum_{i\in U_L}r_i^2
}
where $U_L=(r_{i_1},\cdot\cdot ,{r_{i_L}})$  is any subset of the radial 
polar coordinates $r_i$ ($i:1,..,\Bigl[ {{D-1}\over 2}\Bigr]$).

It is convenient to remark that the MP metric does not depend on the time 
$t$ nor the polar angles $\theta_i$; it is obviously invariant 
under time translations and $\theta_i$ translations. The Killing 
field associated to the first isometry is:
\eqn\tk{
\xi^t ={\partial\over {\partial t}}
}
and the associated conserved charge is the mass. On the other hand, 
the Killing field associated to translations on $\theta_i$ is:
\eqn\tkb{
\xi^i ={\partial\over {\partial {\theta_i}}}
}
and the associated conserved charges are the angular momenta due to 
the rotations along the corresponding planes. These isometries are 
also isometries of the rotated metric \tr , because it is still independent
of the time $t$ and the polar angles $\theta_i$. They play an important 
role when computing the mass, the 
angular momentums and horizon of the general electrically 
charged black-hole.

\newsec{Conserved Charges, Ergosphere and Horizons.}

In this section we are going to discuss general properties of the
electrically charged black-holes that we have obtained in \tr -\tan .
First, we compute the conserved charges of the solution, starting
with the mass and the angular momentum. The appropriate metric to use 
is the Einstein metric, related to \tr\ (from now on we take 
$G_{\mu\nu}=G^{PM}_{\mu\nu}$ in \tr ) by the 
following relation:
\eqn\emp{
{G'}^E_{\mu\nu}=e^{{2\Phi'}\over {2-D}} {G'}_{\mu\nu}.
}

The ADM mass is the charge corresponding to the Killing vector field \tk\ 
and is related to the invariance of the system 
under global time-translations. One obtains:
\eqn\mass{
M={1\over 2}(1+(D-3)\cosh\alpha\cosh\beta)\mu\Omega_{D-2}
}
where $\Omega_{D-2}$ is the area of the unit $(D-2)$-sphere.
The angular momentum $J_i$ is the charge corresponding to the Killing vector 
field \tkb\ and is related to invariance of the solution under global
translations of the polar angle $\theta_i$. The result is:
\eqn\ang{
J_i ={1\over 2}(\cosh\alpha +\cosh\beta)\mu a_i\Omega_{D-2}
}

For the electric charges $Q^{(a)}$ of the black-hole one obtains:
\eqn\cel{
\eqalign{
Q^{(a)}&={\mu\over \sqrt{2}}(D-3)\sinh\alpha\cosh\beta p^a\Omega_{D-2},
\qquad 1\leq a\leq 26-D\cr
Q^{(a+26-D)}&={\mu\over \sqrt{2}}(D-3)\sinh\beta\cosh\alpha q^a
\Omega_{D-2},\qquad 1\leq a\leq 10-D.\cr
}}

Let us now localize the ergosphere. The ergosphere is the 
surface of space-time defined by the equation ${G'}^{E}_{00}=0$.
In our case, from \emp\ and \tr\ we get:
\eqn\erg{
{G'}^{E}_{00}={\Delta}^{{3-D\over {D-2}}}G^{MP}_{00}=0
}
This equation, for finite boosts angles $\alpha$ and $\beta$, 
has the solution $G^{MP}_{00}=0$. Then, the ergosphere of the 
boosted black-holes are at the same place as the ergosphere of the 
MP metric. From \tmp\ we find that the ergosphere is in the region 
where $h=1$, and then, is defined by the equation (see \th ):
\eqn\ergb{
\eqalign{
\Pi F&=\mu{\rho}^2\qquad {\rm for\ D-1\ even}\cr
\Pi F&=\mu\rho \qquad {\rm for\ D-1\ odd}\cr
}}
One can check that all the fields are well defined over 
the ergosphere and that, most notably, the scalar fields take constant values 
there (as one can easily show from \trd\ and \trs ):
\eqn\kes{
\eqalign{
\Phi '|_{erg}&=-\ln \Bigl( {{\cosh\alpha +\cosh\beta}\over 2}\Bigl)\cr
{M'}_{ab} |_{erg}&=\delta_{ab} +\sinh^2\alpha\, p^a p^b,\qquad
1\leq a,b\leq 26-2D\cr
{M'}_{a+26-D,b+26-D}|_{erg}&=\delta_{ab} +\sinh^2\beta\, q^a q^b,\qquad
1\leq a,b\leq 10-D\cr
{M'}_{a,b+26-D} |_{erg}&=0,\qquad 1\leq a\leq 26-d,\quad 1\leq b\leq 10-D.
}}
Also, the temporal component of the gauge fields are constant on 
the ergosphere:
\eqn\gaue{
\eqalign{
\phi^{(a)}&\equiv {A'}^{(a)}_0 |_{erg} =-{1\over\sqrt{2}}
{\sinh\alpha\over{\cosh\alpha +\cosh\beta}}
p^a,\qquad 1\leq a\leq 26-D\cr
\phi^{(a+26-D)}&\equiv {A}^{(a+26-D)}_0 |_{erg}=-{1\over\sqrt{2}}
{\sinh\beta\over{\cosh\alpha +\cosh\beta}}
q^a,\qquad 1\leq a\leq 10-D\cr
}}
Surprisely, we will see below that $\phi^{(j)}$ ($1\leq j\leq
36-2D$) coincides with the electrostatic potential in the event horizon.

Let us now talk about the horizons of the electrically charged 
black-holes. We are going to compute the location of the horizons 
by using the covariant method of Carter \refs{\car}. 
First we construct the ``Killing
form'' (a $\Bigl[ {{D-1}\over 2} \Bigr] +1$ form):
\eqn\car{
K=K_{\mu_0 \mu_1 ,..,\mu_p}dx^{\mu_0}dx^{\mu_1}
\wedge\cdot\cdot\cdot\wedge dx^{\mu_p}
}
where we have defined $p\equiv \Bigl[ {{D-1}\over 2}\Bigr]$ and:
\eqn\carb{
K_{\mu_0 \mu_1 ,..,\mu_p}\equiv {\xi}^{t}_{[\mu_0}
{\xi}^{1}_{\mu_1}\cdot\cdot\cdot {\xi}^{p}_{\mu_p ]}
}
($\xi^{t}$ and $\xi^{i}$ are the Killing fields given in \tk\ and
\tkb ). The horizon is then located at the hypersurface in space-time 
where the norm of this form vanishes ($|K|^2=0$). Let us denote by 
$|{K'}|^2$ the norm of the Killing form computed with the metric
\tr\ (where $G_{\mu\nu}=G^{MP}_{\mu\nu}$) and by $|K^{MP}|^2$ 
the norm computed with the MP metric. 
From \tr\ one finds:
\eqn\mar{
|K'|^2 ={|K^{MP}|^2\over \Delta}
}
As a result, for finite rotation angles $\alpha$ and $\beta$, 
the horizon of the electrically charged black-holes are localized
in the same place as the horizon of the MP metric (however, the area of the
horizon changes due to the change of the metric). The solution to 
the equation $|K^{MP}|^2=0$ is given by:
\eqn\pa{
\eqalign{
\Pi& =\mu\rho \qquad{\rm for\ D\ even}\cr
\Pi& =\mu{\rho}^2 \qquad {\rm for\ D\ odd}\cr
}}
These equations has been studied by Myers and Perry  
(\refs{\myp}). Looking at the form of $\Pi$ \tb\ we see that \pa , 
for $D$ even, is a polynomial equation of degree $D-2$. This means 
that, if they gave solution(s), one can write an explicit formula for 
them in the cases $D=4,6$ \footnote{*}{Here we recall briefly the 
Galois' theorem: the roots 
of a generic polynomial in $\rho$ of order 
$n$ are soluble in terms of radical expressions only for 
$n=2,3,4$.}. For $D$ odd \pa\ is 
a polynomial of degree ${D-1}\over 2$ in ${\rho}^2$. Therefore, the
solution(s) (if any) can also be written explicitly    
for $D=5,7,9$. The cases $D=8,10$ have to be studied separately. 

In \refs{\myp} it is shown that \pa\ can have, at most, two 
solutions $\rho^{\pm}_H$ leading to the inner horizon $\rho^{-}_H$ and the
event $\rho^{+}_H$ horizon. The degenerate case with only one 
solution ($\rho^{+}_H \to\rho^{-}_H$) gives the 
extremal black-hole solution. Finally, the cases 
with no solution produce naked singularities. 

We now wish to compute the physical quantities relevant in 
black-hole thermodynamics. First, we compute the 
null generator ${\tilde k}'$ of the event horizon of the charged
black-holes
(from now on primes denotes quantities associated to the boosted 
black-hole whereas unprimed denote those associated to the 
unboosted MP metric).
${\tilde k}'$ is the combination of Killing fields \tk\ and \tkb\ that 
becomes null on the event horizon. Then, we define:
\eqn\k{
{\tilde k}'=\xi^t -\sum_i {\Omega '}_i \xi^i ,\qquad 1\leq i\leq \Bigl[ {{D-1}
\over 2}\Bigr].
}
Demanding on the event horizon that $|{\tilde k}'|^2 =
{G'^E}_{\mu\nu}{\tilde k}'^{\mu}{\tilde k}'^{\nu}=0$ 
(where ${G'^{E}}_{\mu\nu}$ is the metric given in \emp ) we get:
\eqn\op{
{\Omega '}_i =\bigl( {2\over {\cosh\alpha +\cosh\beta}}\bigr) 
{a_i\over {{\rho^{+}_H}^2 +a^2_i}}
}
To obtain the previous result we have used the fact that, 
on the event horizon, 
from \th\ and \pa\ we have:
\eqn\ut{
h |_{\rho^{+}_H} ={1\over F}
}

The electrostatic potential on the surface of the horizon is given 
by:
\eqn\ep{
\phi^{(a)}=A'^{(a)}_{\mu}{\tilde K}'^{\mu} |_{{\rho}^{+}_H}.
}
Notably, from \trg , \k\ and \op\ 
(and using again \ut )one finds that the electrostatic 
potential on the horizon (which is constant) 
is precisely the temporal component of the 
gauge fields on the ergosphere (as we anticipated in \gaue ).
This seems to be a general rule common to the solutions
constructed using the generating technique, 
and we have checked that the same coincidence happens for the 
Kaluza-Klein black holes of \refs{\horn}.
Next, we compute the surface gravity $\kappa '$ of the rotated black-holes.
$\kappa '$ is defined through the equation:
\eqn\kap{
{\tilde k}'^{\mu}{\nabla '}_{\mu} {\tilde k}'^{\nu}=\kappa ' 
{\tilde k}'^{\nu}
}
evaluated on the horizon ($\nabla '$ is the covariant derivative 
with respect to the metric
\emp ). After a long (but trivial) computation we get 
a simple answer:
\eqn\kapb{
\kappa '={2\over {\cosh\alpha +\cosh\beta}} \kappa
}
where $\kappa$ is the surface gravity of the rotating black-holes 
without electric charges:
\eqn\kapc{
\eqalign{
\kappa &={{\partial_{\rho}\Pi -2\mu\rho}\over {2\mu\rho^2}}|_{\rho^{+}_H},
\qquad {\rm for\ odd\ D}\cr
\kappa &={{\partial_{\rho}\Pi -\mu}\over {2\mu\rho}}|_{\rho^{+}_H},
\qquad {\rm for\ even\ D}\cr
}}
Of course, $\rho^{+}_H$ is a constant on the event horizon and, 
as a result, $\kappa$ is also a constant there.

Finally, the area of the horizon is given by:
\eqn\area{
{A'}_H = (\cosh\alpha +\cosh\beta )
{\mu\over 4\kappa}\bigl( D-3-2\sum_i {a_i\over 
{{\rho^{+}_H}^2+a_i^2}}\bigr)\Omega_{D-2}.
}
The most efficient way of computing the area is by noting that 
$\kappa A_H$ (as is the case for $\sum_i \Omega_i J_i$) is an invariant under 
the generating boosts 
(this is also true for the Kaluza-Klein black holes of 
\refs{\horn}). Then, using 
the result of Myers and Perry for the area of the unboosted case
we arrive at \area .

\bigskip

The case of $\bigl[ {{D-2}\over 2}\bigr]$ rotating planes and two 
electric charges was discussed in \refs{\cvep} (which appeared 
after completion of this work).

\bigskip
{\bf Acknowledgements}

We would like to thank R. Emparan, G.T. Horowitz and D. Marolf for valuable 
discussions. This work is partly supported by a Spanish Government 
grant and partially supported by DOE-91ER40618 and DGICYT (PB93-0344).

\listrefs
\end